\documentclass[aps,amsmath,amssymb,prl,twocolumn,showpacs,superscriptaddress]{revtex4}
\usepackage{graphicx}

\begin{document}

\title{Critical Scale-invariance in Healthy Human Heart Rate}

\author{Ken Kiyono}
\affiliation{Educational Physiology Laboratory, Graduate School of
  Education, The University of Tokyo, 7--3--1 Hongo, Bunkyo-ku, Tokyo
  113--0033, Japan}
\author{Zbigniew R. Struzik}
\affiliation{Educational Physiology Laboratory, Graduate School of
  Education, The University of Tokyo, 7--3--1 Hongo, Bunkyo-ku, Tokyo
  113--0033, Japan}
\affiliation{PRESTO, Japan Science and Technology Agency, Kawaguchi,
  Saitama 332--0012, Japan}
\author{Naoko Aoyagi}
\affiliation{Educational Physiology Laboratory, Graduate School of
  Education, The University of Tokyo, 7--3--1 Hongo, Bunkyo-ku, Tokyo
  113--0033, Japan}
\author{Seiichiro Sakata}
\author{Junichiro Hayano}
\affiliation{Core Laboratory, Nagoya City University Graduate School
  of Medical Sciences, 1 Kawasumi, Mizuho-cho, Mizuho-ku, Nagoya
  467--8601, Japan}
\author{Yoshiharu Yamamoto}
\email{yamamoto@p.u-tokyo.ac.jp}
\affiliation{Educational Physiology Laboratory, Graduate School of
  Education, The University of Tokyo, 7--3--1 Hongo, Bunkyo-ku, Tokyo
  113--0033, Japan}
\affiliation{PRESTO, Japan Science and Technology Agency, Kawaguchi,
  Saitama 332--0012, Japan}

\date{\today}

\begin{abstract}
We demonstrate the robust scale-invariance in the probability
density function (PDF) of detrended healthy human heart rate
increments, which is preserved not only in a quiescent condition,
but also in a dynamic state where the mean level of heart rate is
dramatically changing. This scale-independent and fractal
structure is markedly different from the scale-dependent PDF
evolution observed in a turbulent-like, cascade heart rate model.
These results strongly support the view that healthy human heart
rate is controlled to converge continually to a critical state.
\end{abstract}

\maketitle

Healthy human heart rate belongs to a special class of complex
signals, showing long-range temporal correlations
\cite{1357,1430}, non-Gaussianity of the increment's PDF
\cite{1357} and multifractal scaling properties \cite{1620,Amaral01}, 
and has served as the `benchmark' of choice for studies of biological
complexity. Two alternative mechanisms, both characterized by these
three features, have been proposed for this heart rate complexity:
1) a random (multiplicative) cascade process based on the resemblance
of the behavior of the structure function \cite{Frisch} of heart rate
increments to that of spatial velocity differences in hydrodynamic
turbulence \cite{1759}; and 2) critical state-like dynamics \cite{1522}
based on the resemblance of the scale-invariant properties in heart
rate to those of many systems operating near the critical point of
their phase space. To date the exact mechanism for the complex heart
rate dynamics is unknown. However, as it reflects the dynamics of the
autonomic nervous system's control of heart rate \cite{1357,Amaral01}
and thus provides potential predictors for the mortality of cardiac
patients \cite{1532,1705,1305}, elucidating this mechanism is
considered important.

Lin and Hughson \cite{1759} recently reported an analogy between
turbulence and human heart rate dynamics by finding similarity of the
structure function---directly linked with multifractal formalism
\cite{Muzy93a}---of heart rate increments to that of spatial velocity
differences in a random cascade process proposed as a model of
hydrodynamic turbulence. One of the common features of such
cascade-type multifractal models is the evolution in the shape of the
PDF of the increments from Gaussian at large scales to stretched
exponential at smaller scales \cite{Castaing90}. In this study, we
disprove this cascade-like assertion and demonstrate that heart rate
signals do not follow the evolution in the shape of the (increment)
PDF characteristic for cascade-like processes, but report for the
first time a {\em robust\/} scale invariance. As long-range correlation,
non-Gaussianity and multifractality are also typical characteristics
of a system at the critical point \cite{Hinrichsen02,Kadanoff89},
and fluctuations in a system at a critical point are generally
associated with the scale invariance and universal behavior of the
scaling function \cite{Stanley99,Sethna2001}, we conclude that such
robust scale invariance in the increment PDF suggests the alternative
scenario of the near critical state-like operation for the healthy
heart rate dynamics.

The long-term heart rate data analyzed have been measured as
sequential heart interbeat intervals $b(i)$, where $i$ is the beat
number. We investigate the PDF of heart rate increments at
different time scales (in beat numbers), where the
non-stationarity of the data has been eliminated by local
detrending \cite{1357}. We first integrate the $b(i)$,
$B(m)=\sum_{j=1}^{m} b(j)$, and the resultant $B(m)$ is divided
into sliding segments of size $2n$. Then in each segment
the best $q$-th order polynomial is fit to the data. The
differences $\Delta_s B(i) = B^{*}(i+s) - B^{*}(i)$ at a scale
$s$ are obtained by sliding in time over the segments,
where $B^{*}(i)$ is a deviation from the polynomial fit. By this
procedure, the $(q-1)$-th order polynomial trends are eliminated
and we analyze the whole PDF of $\Delta_s B(i)$.

\begin{figure}[tb]
\begin{center}
 \includegraphics[width = 0.92\linewidth]{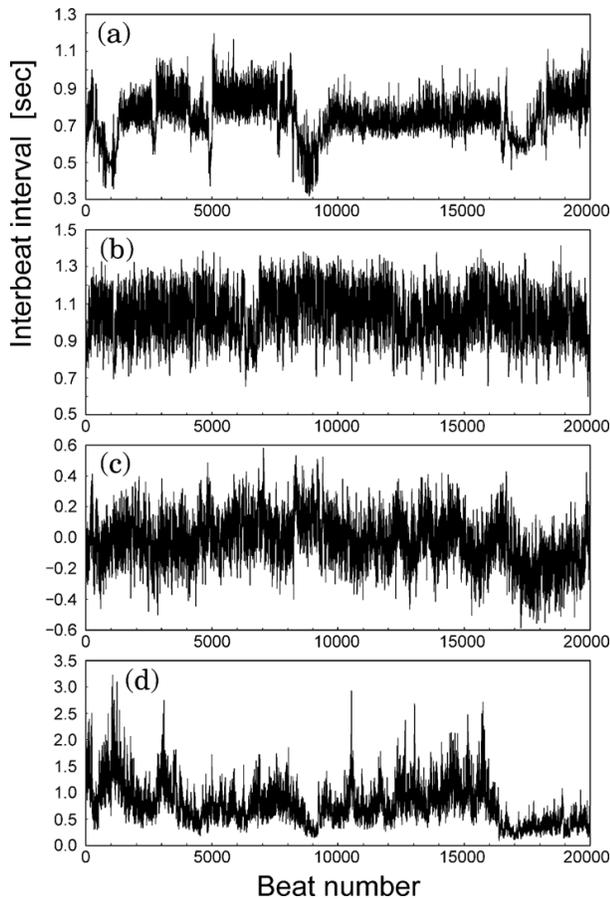}
\end{center}
\caption{
(a) A representative record of daytime (12:00--18:00 hrs) heart
interbeat intervals for a healthy subject. (b) An example of heartbeat 
intervals for a healthy subject during constant routine protocol
\cite{Amaral01,Aoyagi03}. (c) The surrogate data for (b).
(d) Data generated by a cascade heart rate increment model
\cite{1759}. The parameters used for simulation are $J=15$,
$R_t = 2$ and $\sigma_j = R_t^{-2.5-j/J}$ (see Ref.~\cite{1759}
for details). 
}
\label{fig:fig1}
\end{figure}

Using this method, we analyze two experimental and two
computer-generated data sets. The first data set consists of
daytime (12:00--18:00 hrs) heart rate data with a length of up to
$4\times 10^4$ heartbeats from 50 healthy subjects (10 females
and 40 males; ages $21-76$ years) without any known disease
affecting the autonomic control of heart rate [Fig.~\ref{fig:fig1}(a)].
Details of the recruitment of the subjects, screening for medical
problems, protocols and the data collection are described in
Sakata {\it et al.\/} \cite{Sakata99}. The data were collected
during normal daily life. The second experimental data set
consists of data of seven 26-hour long periods (up to $10^5$ beats),
collected when the subjects (7 males; ages $21-30$ years) underwent
`constant routine' (CR) protocol, where known behavioral factors
affecting heart rate (e.g. exercise, diet, postural changes and sleep)
are eliminated [Fig.~\ref{fig:fig1}(b)] \cite{Amaral01,Aoyagi03}.

In order to test the possible presence of non-linear mechanisms
in complex heart rate dynamics, we apply the surrogate data test
to the CR protocol data  \cite{Schreiber00}. We generate a
surrogate data set with the same Fourier amplitudes and 
distributions as the original increments in the CR protocol data
[Fig.~\ref{fig:fig1}(c)]. Since only linear temporal correlation
of $b(i+1)-b(i)$ is retained in the surrogate data, a comparison
with the raw data can be used to test whether the PDF of
`velocity' increments $\Delta_s B(i)$ possesses some non-linear
mechanism inherent to it. Finally, we generate heart rate
increments of comparable lengths, following the `turbulence-like'
scenario from the random cascade model proposed recently by Lin
and Hughson [Fig.~\ref{fig:fig1}(d)] \cite{1759}.

\begin{figure*}[tb]
\begin{center}
 \includegraphics[width = 0.82\linewidth]{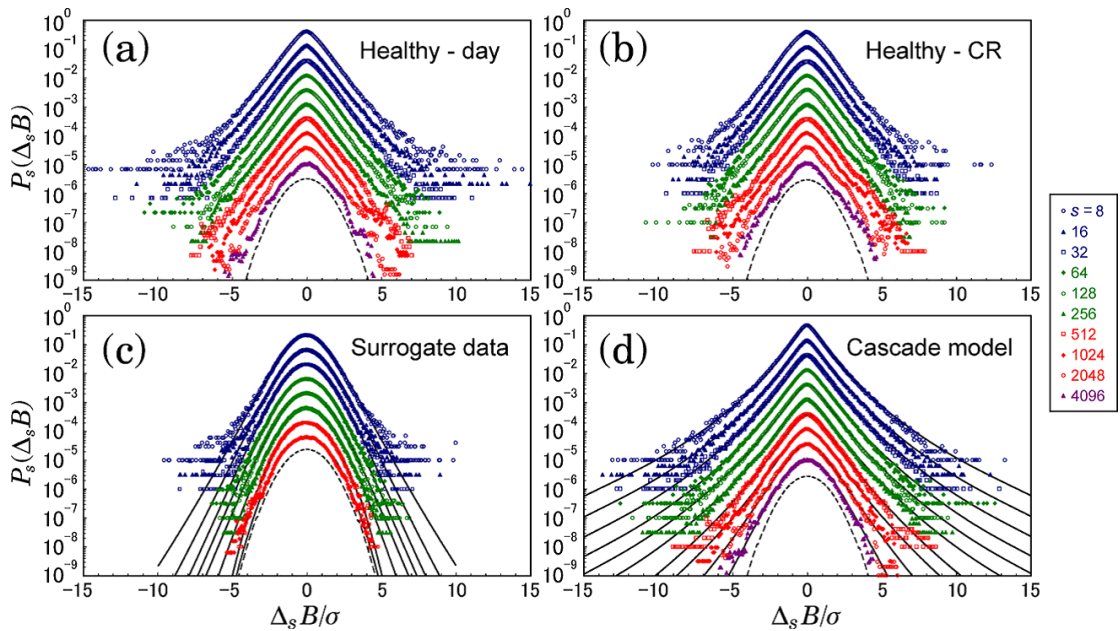}
\end{center}
\caption{
Deformation of increment PDF's across scales. Standardized PDF's
(in logarithmic scale) of $\Delta_s B(i)$ for different time
scales are shown for (from top to bottom) $s$ = 8, 16, 32, 64,
128, 256, 512, 1024 beats. These PDF's are estimated from all
samples in each group. The dashed line is a Gaussian PDF for
comparison. (a) The PDF's from daytime (12:00--18:00 hrs) heart
rate time series from healthy subjects. (b) From healthy
subjects during constant routine protocol. (c) From surrogate
time series for (b). (d) From a cascade model. Note
in the last case (d) the continuous deformation and the
appearance of fat tails when going from large to fine scales.
In solid lines, we superimpose the deformation of the PDF using
Castaing's equation with the log-normal self-similarity kernel,
providing an excellent fit to the data.
}
\label{fig:fig2}
\end{figure*}

PDF's of $\Delta_s B(i)$ for healthy humans, which are standardized by
dividing the heart rate increments in each record by the standard
deviation, are non-Gaussian in shape for a wide range of scales
$8 \leq s \leq 4,096$ irrespective of whether the subjects were
in their normal daily routine [Fig.~\ref{fig:fig2}(a)] or in CR
[Fig.~\ref{fig:fig2}(b)]. In contrast, the PDF's of the surrogate
data are near Gaussian, although non-Gaussianity with the fat tails
close to those in the observed data is still encountered at fine
scales [Fig.~\ref{fig:fig2}(c)]. The difference between the healthy
human and surrogate data indicates that the observed non-Gaussian
behavior is related to non-linear features of the healthy heart
rate dynamics. The PDF's of the cascade model show continuous
deformation and the appearance of fat tails when going from large
to small scales [Fig.~\ref{fig:fig2}(d)]. 

\begin{figure*}[tb]
\begin{center}
 \includegraphics[width = 0.70\linewidth]{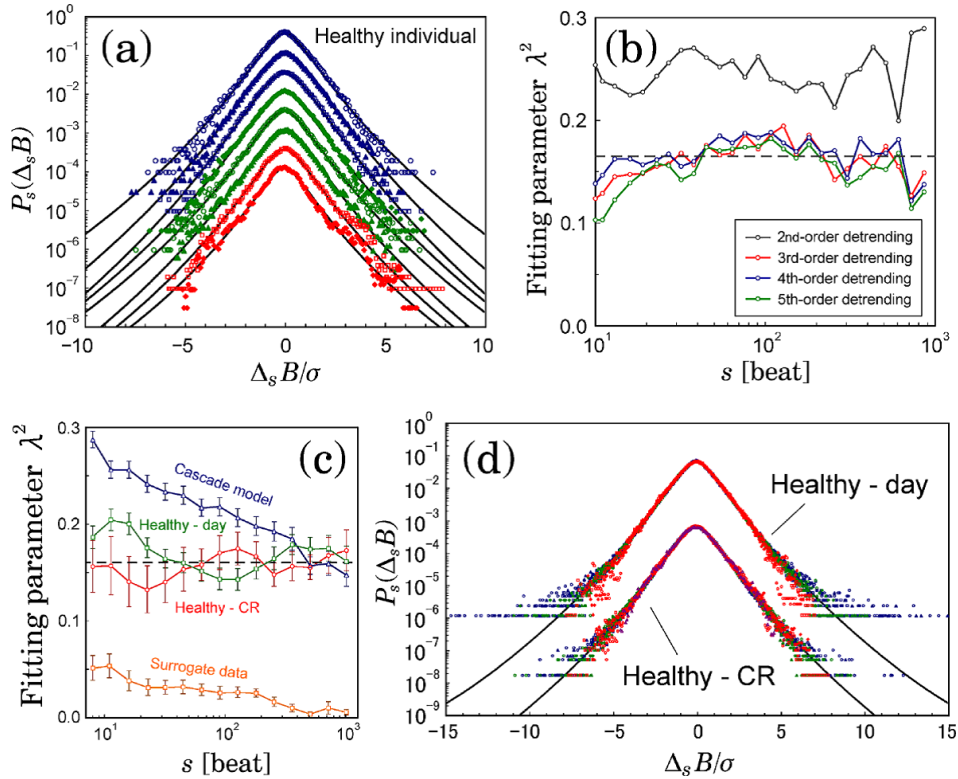}
\end{center}
\caption{
(a) Standardized PDF for a single subject during constant
routine protocol [see Fig.~\ref{fig:fig1}(b)]. In solid lines,
we superimpose the PDF's using Castaing's equation. (b)
Dependence of the fitting parameter $\lambda^2$ of Castaing's
equation on the order of detrending polynomials. (c)
Dependence of the fitting parameter $\lambda^2$ on the scale $s$.
The error bars indicate the standard error of the group averages.
(d) Superposition of standardized PDF's at different scales
shown in Fig.~\ref{fig:fig2}a and \ref{fig:fig2}b, where we use
the scale range $8 \leq s \leq 2048$ and $8 \leq s \leq 4096$,
respectively. In solid lines, we superimpose the PDF using the
Castaing's equation ($\lambda^2 = 0.16$ for both groups).
}
\label{fig:fig3}
\end{figure*}

For a quantitative comparison, we fit the data to the following 
function based on Castaing's equation \cite{Castaing90}: 
\[
\tilde{P}_s(x) = \int P_{L} \left( \frac{x}{\sigma} \right)\frac{1}{\sigma}
G_{s,L}(\ln \sigma) {\rm d} (\ln \sigma), 
\]
where $P_{L}$ is the increment PDF at a large scale $L>s$, and
the self-similarity kernel $G_{s,L}$ determines the nature of
the cascade-type multiplicative process. Here we assume $P_{L}$
and $G_{s,L}$ are both Gaussian, 
\[
G_{s,L}(\ln \sigma) = \frac{1}{\sqrt{2 \pi} \lambda}
\exp \left(- \frac{\ln ^2 \sigma}{2 \lambda^2} \right), 
\]
and investigate the scale dependence of $\lambda^2$. The fit of
the PDF of actual heart rate increments to Castaing's equation
is indeed almost perfect, especially within $\pm 3$ times standard
deviation, even for a single record [Fig.~\ref{fig:fig3}(a)], and
robust in terms of the effect of the order of detrending
polynomials on the estimation of $\lambda^2$, if the order is
greater than two [Fig.~\ref{fig:fig3}(b)]. In the following, we use
the third order detrending for the estimation of $\lambda^2$.

Within the turbulent cascade picture, the parameter $\lambda^2$
can be interpreted as being proportional to the number of cascade
steps and is known to decrease linearly with $\log s$
\cite{Castaing90,Ghashghaie96}. The cascade heart rate model
studied here \cite{1759} also shows this effect [Fig.~\ref{fig:fig3}(c)].
In contrast, the scale dependency of $\lambda^2$ for healthy heart
rate increments is remarkably different [Fig.~\ref{fig:fig3}(c)].  
Especially during CR, we cannot see any decrease in $\lambda^2$
with $\log s$. There is no significant difference in the average
$\lambda^2$ at different scales, tested by the analysis of variance,
over the range of 23--2,048 beats for healthy subjects during daily
routine and 8--4,096 beats during CR [$F(13,686)=1.73$ and
$F(18,114)=1.70$, respectively, $p>0.05$], and the slopes of
$\lambda^2$ vs. $\log s$ are much closer to zero, which means the
absence of cascade steps across the scales in the corresponding range.  

In addition, when the PDF's at different scales are superimposed
[Fig.~\ref{fig:fig3}(d)], all the data collapse on the same curve,
which is one of the characteristic features observed in fluctuations
at a critical point \cite{Mehta02}.
The range of scales where this scale-invariance of PDF is observed,
spanning from about $10$ beats to a few thousand heartbeats, is
compatible with that of the robust, behavioral-independent $1/f$
scaling \cite{Aoyagi03} and multifractality \cite{Amaral01} of
heart rate. The scale-invariance in the PDF is also robust in the
sense that it is observed not only during CR but also during normal
daily life, where behavioral modifiers of heart rate 
dramatically change the mean level of heart rate [e.g.
Fig.~\ref{fig:fig1}(a)]. We thus find a novel property of scale
invariance in healthy human heart rate dynamics, reminiscent of
systems in a critical state. In particular, the invariance discovered
strongly supports the view that the healthy human heart is controlled
to converge continually to a critical state. Such a critical point
itself may, however, be shifted by the effects of the external and/or
internal environment. 

Struzik {\it et al.\/} \cite{Struzik04} recently demonstrated that
modifying the relative importance of either the sympathetic or
the parasympathetic branch of the autonomic nervous system leads
to a substantial decrease in $1/f$ scaling, showing that $1/f$
scaling in healthy heart rate requires the existence of and the
intricate balance between the antagonistic activity of these two
branches. They further suggest the view of cardiac
neuroregulation as a system in a critical state \cite{Bak87}, and
permanently out of equilibrium, in which concerted interplay of
the sympathetic and parasympathetic nervous systems is required
for preserving momentary `balance'. Our findings provide more
direct evidence for this. The precise mechanism
responsible for {\it critical\/} heart rate dynamics requires
further research.  It is of note, however, that there exists a 
physiological model for the dynamics of cardiac neuroregulation
\cite{Kotani02}, equipped with antagonistic and multiplicative
delayed feedback loops, within time scales where the critical
scale-invariance in heart rate is observed. 
The mechanism of the critical mode of operation could be clarified 
by investigating essential dynamics in such a `first principles', 
non-linear physiological model.

The functional advantage of the heart rate control system being in a
critical state remains an open question.  However, an analogy with
other critical phenomena might help to understand this.
Studies on transport properties through
complex networks \cite{Takayasu00,Valverdea02} have demonstrated
maximum efficiency of transportation at the critical point, which
is the  phase transition point from an `uncrowded' state to a
`congested' state in the transportation routes. % In the case of the
% heart, these would correspond typically to a state with central
% hypovolemia (uncrowded) and that with `congestive' heart failure
% (congested). Both cases are known to be associated with decreased
% cardiac output due to a lack of blood to pump out and difficulty
% in pumping out the blood, respectively \cite{Guyton}.
Thus, our results may indicate that the central neuroregulation continually
brings the heart to a critical state to maximize its functional
ability, coping with the continually changing pre- and after-load
on the heart. This may be particularly important in understanding
the widely reported evidence that decreased $1/f$ variability
\cite{1532,1705}, especially in the low frequency region \cite{1532,1305},
is associated with increased mortality in cardiac patients.
To date there have been no successful attempts to provide a
satisfactory explanation for this. We suggest for the first time
that a breakdown of the optimal control achieved in the critical
state might be related to this clinically important phenomenon.

This work was in part supported by grants from the Research Fellowships
of the Japan Society for the Promotion of Science for Young Scientists
(to K.K.) and from Japan Science and Technology Agency (to Y.Y.).

\end{document}